\documentclass[showpacs,epsf,twocolumn]{revtex4-1}
\usepackage{float}
\usepackage{color,graphicx}
\usepackage{hyperref}
\usepackage{amsmath}
\begin{document}

\title{Correlation between battery material performance and  cooperative electron-phonon interaction in LiCo$_y$Mn$_{2-y}$O$_{4}$}
\author{Krishna Rao Ragavendran, Prabhat Mandal, and Sudhakar Yarlagadda}
\affiliation{Saha Institute of Nuclear Physics, HBNI, 1/AF Bidhannagar, Kolkata 700 064, India}
\date{\today}

\begin{abstract}
{Understanding the basic physics related to archetypal lithium battery material
(such as LiCo$_y$Mn$_{2-y}$O$_{4}$) is of considerable interest and is expected to  aid designing
of cathodes of high capacity.
The relation between electrochemical performance, activated-transport parameters, thermal expansion, and cooperativity of
electron-phonon-interaction distortions in LiCo$_y$Mn$_{2-y}$O$_{4}$ is investigated.
The first order cooperative-normal-mode transition,   detected through coefficient of thermal expansion, is found to disappear
at a critical doping ($y \sim 0.16$);  interestingly, for $y \gtrsim 0.16$ the resistivity does not change much with doping and the electrochemical
capacity becomes constant over repeated cycling.
The  critical doping $y \sim 0.16$ results in breakdown of the network of  cooperative/{coherent}
normal-mode distortions; this leads to vanishing of the first-order transition, establishment of hopping
channels with lower resistance, and enhancing lithiation and delithiation of the battery, thereby minimizing electrochemical capacity fading. }
\end{abstract}
\maketitle

\section{Introduction}
{Among the commercially available batteries, lithium-ion batteries have the highest energy density
and are the primary energy source for portable electronics.
The energy stored in rechargeable lithium batteries is limited by their
cathodes.
Currently, cutting-edge cathodes use
either layered LiCoO$_2$, its three-dimensional doped variant LiM$_y$Mn$_{2-y}$O$_{4}$ (with M=Ni,Co, etc.),
or polyanionic compounds such as olivine-type LiFePO$_4$ \cite{review0,review1,review2,review3,review4}}.
Introduced in 1983 by J. B. Goodenough \cite{good}, Li$_x$Mn$_{2}$O$_{4}$ serves as an attractive alternative to Li$_x$CoO$_2$ in terms of environmental friendliness, cost effectiveness,  safety \cite{wakihara},
as well as with its virtue to enable the variation of the lithium stoichiometry ‘$x$’ over the entire range
from 0 to 1. When $x=0$,  the battery is said to be in a completely charged state, while $x=1$ implies the battery
is totally discharged.   The material, however, suffers from the problem of electrochemical capacity loss over repeated cycling, 
which is circumvented by doping at the Mn site with other transition metal atoms \cite{wakihara2}.
By a trial and error approach, it is established that among the doped variants, the cobalt doped spinel
LiCo$_y$Mn$_{2-y}$O$_4$ with doping level between 0.1 and 0.2 is ideal for
enhanced battery performance
\cite {wu, huang, arora, shen, amdouni}.
However, to the best of our knowledge, no clear correlation has been made between battery performance
and basic physics related phenomena such as structural  phase transitions, anomalies in
transport and heat capacity, etc.
Thus, a detailed understanding of the fundamental physics of LiCo$_y$Mn$_{2-y}$O$_4$ is  highly essential for
developing cutting-edge energy storage devices.

Phase transitions in LiMn$_{2}$O$_{4}$ (and its doped variants)
can be grouped under two categories on the basis of the temperature at which the transitions occur: the high temperature region (200 K to 300 K) involving coherent normal-mode distortions \cite{good2}
(and charge ordering \cite{car}), and the low temperature region ($\sim$ 60 K) corresponding to the magnetic (i.e, antiferromagnetic) ordering. Phase transitions in the battery material in the former region
 (which is close to the room temperature) pertains  to a structural transition, is expected to be
of significance to the battery performance, and will be studied in this work.

{
{The MnO$_6$ octahedra in spinel type LiMn$_2$O$_4$ are similar
 to the MnO$_6$ octahedra in perovskite manganites (such as LaMnO$_3$);
 hence, the framework of coherent normal modes [such as Jahn-Teller (JT) modes]
 utilized in perovskite manganites is expected to provide insights
 into normal-mode ordering in the spinel structures.}}
 Using experiments and theory, we try
to understand the sophisticated physics behind the dopant level of LiCo$_y$Mn$_{2-y}$O$_{4}$ in the stabilization of cathode materials for
energy device applications, so that we can design new materials to transcend the
existing performance levels.

In this paper, we investigate transport,  structural, and thermal properties of both
pure and doped LiCo$_y$Mn$_{2-y}$O$_{4}$ compounds and examine the effect of these properties on battery performance.
We explain the superior electrochemical performance of cobalt-doped LiMn$_{2}$O$_{4}$
within the framework of cooperative-normal-mode physics.
When  non-JT ions such as Co$^{3+}$
replace the JT Mn$^{3+}$ ions, the cooperative link connecting the local normal-mode distortions starts to weaken and the electronic conductivity increases due to tunneling of charge carriers through regions of non-cooperative normal modes.  As the level of Co reaches a certain threshold (i.e., $y \sim 0.16$), the cooperative link is completely broken and the electrochemical discharge capacity shows stability with the cycle number. Any further increase in the cobalt level, while ensuring a stable discharge capacity, only contributes to a lowering of the capacity
due to a decrease in the number of carriers; this explains the emergence of $y \sim 0.16$ as the ideal doping level
for battery performance.

\section{Experimental procedures}
{
{Spinel type polycrystalline LiCo$_y$Mn$_{2-y}$O$_4$ samples with different values of ‘$y$’
were prepared through a conventional solid state method. Stoichiometric quantities of Li$_2$CO$_3$,
MnCO$_3$ and CoCO$_3$ were ground and calcined in platinum crucibles at $800\,^{\circ}\mathrm{C}$ for 
16 h, with intermittent grindings, in an atmosphere of air. The samples were then pelletized and sintered
at $800\,^{\circ}\mathrm{C}$ for 16 h. The phase purity of  LiCo$_y$Mn$_{2-y}$O$_4$ samples 
was checked by high resolution powder x-ray diffraction (XRD) method with Cu K$_{\alpha}$ radiation in 
a Rigaku x-ray diffractometer (TTRAX II). Within the resolution of XRD, we have not observed any peak
due to  impurity phase(s). The XRD pattern can be indexed well with 
the space 
group symmetry  Fd3m. Four probe measurements of the electrical resistivity were done in a commercial 
cryostat (Cryogenic Ltd.) using Keithely 6514 Electrometer. For the thermal expansion measurement, 
a miniature tilted plate capacitance dilatometer was used. The capacitance has been measured by 
an Andeen Hagerling 2700A Ultra-precision Capacity Bridge. In this technique, a change in the sample
length, ${\rm \Delta L=L(T)-L_0}$, where ${\rm L_0}$ is the length of the sample at the lowest measured
temperature, can be determined very accurately. We have measured the macroscopic length change 
of a sample of dimension 1 mm  and the coefficient of linear thermal expansion [$\alpha(T)$] has
been calculated using the relation, $\alpha(T)={\rm \frac{1}{L_0}\frac{d}{dT}\Delta L}$. }\\

\section{theory, results, and discussion}
The Hamiltonian for the cooperative electron-phonon interaction system contains the hopping term $H_t$,  the
electron-phonon interaction  $H_{ep}$,
and the lattice term $H_l$. Here, in the $H_{ep}$ term,
we consider only the breathing mode (BM) and the JT modes;
rest of the normal modes are not taken into account.
At the site $(i,j,k)$, we define the creation operators for $d_{x^2-y^2}$  and $d_{z^2}$ orbitals
as $ d^\dagger_{x^2-y^2;i,j,k}$ and $ d^\dagger_{z^2;i,j,k} $, respectively.
The  indices $i$, $j$, and $k$
correspond to labels along the $x$-, $y$-, and $z$-axes, respectively.
 Next, we express $H_{ep}$
 in  the  orthogonal basis $\psi_{x^2-y^2}$ and $\psi_{z^2}$ as follows \cite{el-ph,allen}:
\begin{widetext}
\begin{eqnarray}
H_{ep} = &-&\frac{1}{4} g \omega_0 \sqrt{2 M \omega_0} \nonumber \\
&\times&\sum_{i,j,k} (d^\dagger_{z^2;i,j,k},d^\dagger_{x^2-y^2;i,j,k})
\begin{pmatrix}
Q_{x;i,j,k}+Q_{y;i,j,k}+4Q_{z;i,j,k} & -\sqrt{3}Q_{x;i,j,k}+\sqrt{3}Q_{y;i,j,k} \\
-\sqrt{3}Q_{x;i,j,k}+\sqrt{3}Q_{y;i,j,k} & 3Q_{x;i,j,k}+3Q_{y;i,j,k}
\end{pmatrix}
\begin{pmatrix}
d_{z^2;i,j,k} \\
d_{x^2-y^2;i,j,k}
\end{pmatrix} ,
\label{eq:gen_elph_com}
\end{eqnarray}
\end{widetext}
where $g$ is the electron-phonon coupling, $\omega_0$ is the frequency of optical phonons,
$M$ is the mass of an oxygen ion. Furthermore, we define $Q_{x;i,j,k} \equiv  u_{x;i,j,k}-u_{x;i-1,j,k}$
where $u_{x;i,j,k}$ and $u_{x;i-1,j,k}$ are the displacements in the $x$-direction of the
two oxygen ions along the $x$-axis around the site $(i,j,k)$; $Q_{y;i,j,k}$ and $Q_{z;i,j,k}$
are defined  similarly in terms of the displacements in the $y$- and $z$-directions of the
oxygen ions along the $y$- and $z$-axes around the site $(i,j,k)$, respectively.

Now, the breathing mode $Q_1$ has all the Mn-O bond lengths changing
uniformly and is defined as $Q_1 \equiv (Q_x +Q_y+Q_z)\sqrt{2/3}$.
The three-dimensional tetragonal JT
distortion mode, commonly known as the $Q_3$ mode and defined
as $Q_3 \equiv(2Q_z -Q_x -Q_y)/\sqrt{3}$,
has the Mn-O
bond lengths changing in all directions \cite{allen,khomskii}. On the other hand,
the planar JT distortion mode, commonly
known as the $Q_2$ mode, has the Mn-O bond lengths changing only
in the plane and is defined as $Q_2 \equiv Q_x-Q_y$. If a single $e_g$
electron occupies a Mn site, from Eq. (\ref{eq:gen_elph_com}), it is clear
that in the $d_{z^2}$ orbital the electron would excite both $Q_1$ and $Q_3$ modes, whereas
in the $d_{x^2-y^2}$ orbital
it would generate only the planar $Q_2$ excitation.

\begin{figure}[b]
\includegraphics[width=0.5\textwidth]{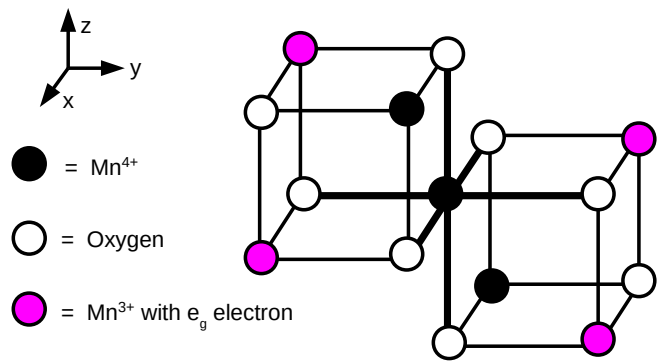}
\caption{Part of the unit cell of LiMn$_{2}$O$_{4}$ in an ideal spinel lattice. Octahedral coordination
to a manganese is depicted using thicker lines. Each manganese tetrahedron in a cube
has two JT-active Mn$^{3+}$ ions and two non-JT Mn$^{4+}$ ions.
}\label{Fig. 5}
\end{figure}

\begin{figure}[b]
\includegraphics[width=0.5\textwidth]{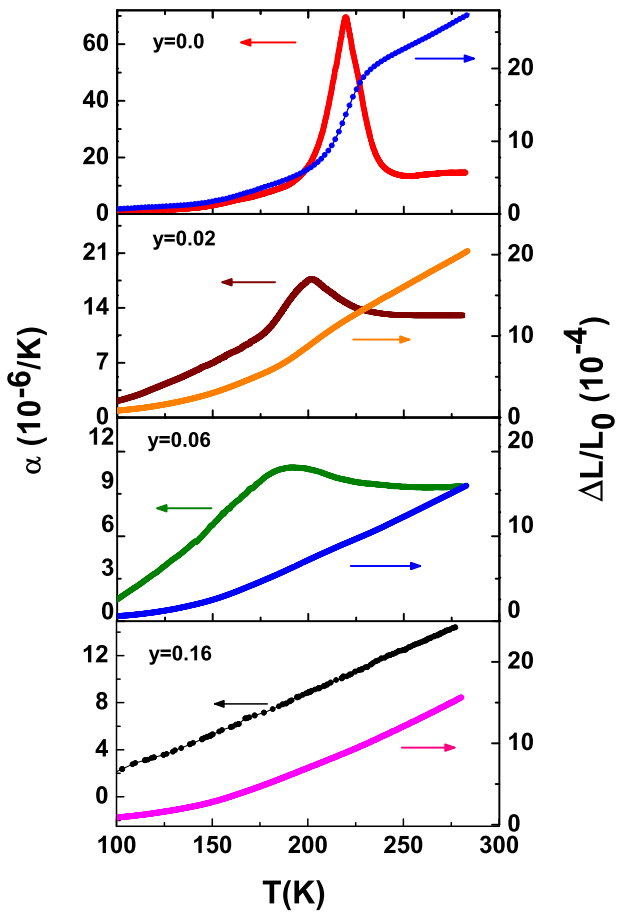}
\caption{Thermal expansion measurements on  cobalt doped LiMn$_2$O$_4$. With increasing doping,
there is a decrease in  the transition temperature, weakening of the peak
in linear thermal expansion $\alpha$($T$), and a drop in the relative expansivity ${\rm \Delta L/L_0}$
 at the structural transition.
 At a critical doping $y \sim 0.16$, the first-order structural transition vanishes
 due to the forging of a percolating path involving non-cooperative normal modes.
}
 \label{Fig. 1}
\end{figure}

We will first consider the undoped LiMn$_2$O$_4$ spinel. As shown in Fig. \ref{Fig. 5},
to minimize energy due to Coulombic interactions,
there are two Mn$^{3+}$ ions and two Mn$^{4+}$ ions in each cube.
Furthermore, from the above equation (\ref{eq:gen_elph_com}) and
Fig. \ref{Fig. 5}, it is clear that when only $d_{x^2-y^2}$ orbital is occupied by
an $e_g$ electron at a
Mn$^{3+}$ ion in the
LiMn$_2$O$_4$
spinel,
there is no frustration and
the energy is minimized. This is because, in any
face that is in the $xy$-plane of a cube
shown in
Fig. \ref{Fig. 5}, we have one Mn$^{3+}$ ion and one Mn$^{4+}$ ion diagonally
opposite to each other. Within the above occupancy scenario,
only planar $Q_2$  mode is excited at Mn$^{3+}$ sites; then, the distortion at one
Mn$^{3+}$ site is compatible with the distortion at another Mn$^{3+}$ site,
thereby cooperatively avoiding frustration.
In addition to the BM and JT modes, if other normal modes are also included in our considerations,
there may be a slight frustration.
{
{In fact, compared to the large JT ordering temperatures
(i.e., at least 750 K \cite{prabhat_oo}) in perovskite manganites, the first-order structural transition
occurs at a much lower temperature (i.e., about 220 K as shown in Fig. \ref{Fig. 1}) indicating a much
weaker frustration due to interaction between cooperative normal-mode distortions.
Furthermore,  the small overall relative expansion around the
structural transition shown in Fig. \ref{Fig. 1} (which is an order of magnitude smaller than in perovskite manganites \cite{prabhat_oo})
also reflects the relevance of, besides the BM and JT modes, additional cooperative electron-phonon distortion mode(s).}}

Next, we will  analyze  the relative expansion ${\rm \Delta L/L_0}$ in the doped system  LiCo$_y$Mn$_{2-y}$O$_4$.
In the undoped LiMn$_2$O$_4$,  a structural transition  occurs as the temperature is lowered
to $T \sim 220$ K (as shown in Fig. \ref{Fig. 1})
and it indicates onset of cooperative normal-mode distortions.
As the doping increases,  a fraction of the JT active Mn$^{3+}$ (shown in Fig. \ref{Fig. 5})
are replaced by  non-JT   Co$^{3+}$ ions \cite{amdouni,wakihara3};
consequently, the frustration decreases and the  cooperative ordering becomes less robust
leading to a decrease in the transition temperature of the first-order structural transition.
At the structural transition, as can be seen from the plots in Fig. \ref{Fig. 1},
there is a sizeable
change in the relative expansivity 
{${\rm \Delta L/L_0}$}; this change in ${\rm \Delta L/L_0}$
diminishes with increasing doping (again illustrating weakening
of cooperative ordering) and finally vanishes at a critical doping $y \sim 0.16$.
At the critical doping $y \sim 0.16$, a percolating path involving
non-cooperative normal-mode distortions is established.
{
{The progressive suppression of the JT transition with  
Co doping is more clearly visible in the temperature dependence of the coefficient
of linear thermal expansion. For undoped sample, $\alpha$($T$) shows a peak around 220 K; 
the peak weakens, becomes broader, and shifts to lower temperatures
as Co concentration increases and disappears at around $y$=0.16.}

{
{This picture of weakening cooperative distortions is similar to that observed
in systems such as perovskite manganites where the JT ordering temperature diminishes with
doping and finally vanishes.  A depiction of JT ordering
temperature versus doping in La$_{1-x}$Ca$_x$MnO$_3$ is given in Refs. \onlinecite{cheong1,tvr}}}.

\begin{figure}[t]
\includegraphics[width=0.7\textwidth]{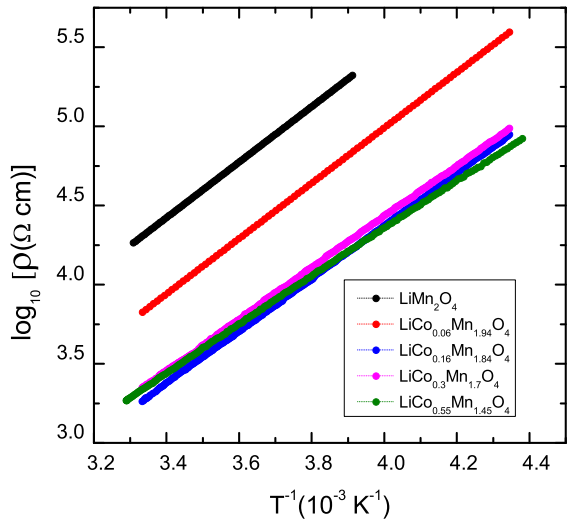}
\caption{Linearity in plots of 
{$\log_{10}(\rho)$} vs ${\rm 1/T}$ depicting activated transport
at various dopings and above the structural transition.
For doping below $y \sim 0.16$,
 the resistivity drops sizeably
with increase in doping; contrastingly, for $y \gtrsim 0.16$, the resistivity does not change
much with doping.
}
\label{Fig. 2}
\end{figure}

{
{
Before, discussing conduction in undoped and doped LiMn$_2$O$_4$,
we will first recapitulate conduction mechanisms in the two-band perovskite manganites
and the single-band Holstein model.
The crossover from hopping conduction to band-like conduction
occurs at lower temperatures in wider-band perovskite manganites
such as La$_{1-x}$Ca$_x$MnO$_3$ and La$_{1-x}$Sr$_x$MnO$_3$ \cite{dagotto}.
Now, 
within the two-band picture of perovskite manganites in Ref. \cite{tvr},
the  upper broad band (due to undistorted
states that are orthogonal to the polaronic states) 
overlap with the polaronic band to produce conduction  at carrier 
concentrations corresponding to $0.2 \lesssim x \lesssim 0.5$.
Here, in the case of LiCo$_y$Mn$_{2-y}$O$_4$, since the
hopping integral is smaller than in manganites \cite{atanasov,seshadri}, the upper band
is not relevant for conduction.
Next, in the single-band Holstein model, band-like conduction occurs
when  band-width is larger than twice the
uncertainty in energy due to electron-phonon scattering (i.e., $~\hbar/\tau$ with $\tau$ being the scattering time)
\cite{ys,holstein}. However, in the present  cobalt-doped LiMn$_2$O$_4$,
the narrow band width does not produce band-like conduction even at low temperatures
and, consequently, only hopping conduction due to a single polaronic band is realized.}

We will now discuss the conduction in undoped LiMn$_2$O$_4$.
Since the $e_g$ electron at a Mn$^{3+}$ site distorts the oxygen cage around it and forms a small
polaron (with polaronic energy $E_p \propto g^2 \omega_0$) \cite{ys,holstein},
the transport of the $e_g$ electron at higher temperatures
(i.e., around room temperature) will be activated with
the activation energy $E_a$ being at least half the polaronic energy $E_p$. In fact,
the value of $\frac{(E_a-E_p/2)}{E_p/2}$, reflects the degree of frustration due to cooperative electron-phonon modes
as will be explained below.
Now, each Mn$^{4+}$ site has four Mn$^{3+}$ sites
diagonally opposite to it as shown in Fig. \ref{Fig. 5}.
If the $e_g$ electron  hops from a Mn$^{3+}$ ion to a neighboring Mn$^{4+}$ ion,
the distortion produced by the remaining three Mn$^{3+}$ ions (which are neighbors of the Mn$^{4+}$ ion)
makes the hopping unfavorable
leading to enhancement in the activation energy $E_a$ by a fraction of the
polaronic energy.

\begin{figure}[t]
\includegraphics[width=0.55\textwidth]{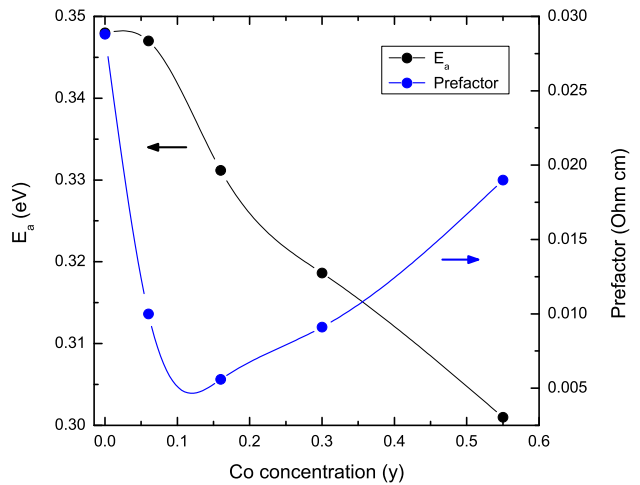}
\caption{Plots of the 
{prefactor} $A \frac{ e^{(2 R/\xi)}}{n}$ and the activation energy $E_a$
[occurring in the resistivity equation (\ref{rho})] as a function
of cobalt doping in  LiCo$_y$Mn$_{2-y}$O$_4$. The prefactor shows a minimum
at $y \sim 0.16$ and the drop in $E_a$ is sharper till $y \sim 0.16$;
both are indicative of a change in the transport mechanism.}\label{Fig. 3}
\end{figure}


 Next, we will focus on the conduction in  doped LiMn$_2$O$_4$.
In LiCo$_y$Mn$_{2-y}$O$_4$, since Co$^{3+}$
replaces Mn$^{3+}$, the corners of each cube in Fig. \ref{Fig. 5} (on an average)  are occupied by $2-2y$  Mn$^{3+}$ ions,
$2y$ Co$^{3+}$ ions, and two Mn$^{4+}$ ions.
Since Co$^{3+}$ is not JT active, the
oxygen-cage  surrounding Co$^{3+}$ is not distorted. Consequently, hopping
from Mn$^{3+}$ to Mn$^{4+}$ has, on an average, a lower  activation energy due to decrease in frustration.
%
The resistivity in  LiCo$_y$Mn$_{2-y}$O$_4$
can be expressed as follows \cite{mott}
\begin{eqnarray}
 \rho = A \frac{ e^{(2 R/\xi)}}{n}e^{(E_a/k_BT)} ,
 \label{rho}
\end{eqnarray}
where
$A$ is a constant,
$n$ is the concentration of the mobile $e_g$ electrons, $\xi$ is the localization length,
$R$ is the shortest hopping distance for an $e_g$ electron
(i.e., the distance between two neighboring Mn$^{3+}$ and Mn$^{4+}$ ions),
and $T$ is the temperature.
The fact that, even upon doping with cobalt,
each Mn$^{3+}$ has  the same number of diagonally opposite Mn$^{4+}$ ions for the $e_g$
electron to hop to, justifies using a fixed-hopping-distance model rather than a
variable-hopping-range model. Our transport model is clearly verified
 by Fig. \ref{Fig. 2} which depicts  linear plots of
$\log_{10}(\rho)$ versus ${\rm 1/T}$. Using Fig. \ref{Fig. 2}, at various Co-doping values, we extract
the prefactor $A\frac{ e^{(2 R/\xi)}}{n}$ and the activation energy $E_a$
[occurring in Eq. (\ref{rho})] and generate Fig. \ref{Fig. 3}. Now, the localization length $\xi$
decreases with the cobalt doping. Hence, in the above  expression for resistivity,
the prefactor $A\frac{ e^{(2R/\xi)}}{n}$
will have a minimum as a function of Co-doping because $\frac{ 1}{n}$ decreases with doping
while the term $ e^{(2 R/\xi)}$
increases with doping. In fact, as depicted in Fig. \ref{Fig. 3}, the minimum in the prefactor
$A\frac{ e^{(2 R/\xi)}}{n}$ occurs at $y \sim 0.16$.
Here it should be pointed out that, below $y \sim 0.16$,
localization length decreases slowly  with doping  (due to cooperative normal-mode-network
weakening with doping); whereas above $y\sim 0.16$, localization length  decreases more rapidly with
doping.
Next, the activation energy $E_a$ (as shown in Fig. \ref{Fig. 3})  monotonically decreases
with doping; however, the drop is sharper till  $y \sim 0.16$. Lastly,
it is of interest to
 note that the resistivity
drops sizeably with increasing doping until the doping-level attains a value $y \sim 0.16$; at higher
doping values (i.e., $y \gtrsim 0.16$), the resistivity does not change much (as can be seen in Fig. \ref{Fig. 2}).
This can be understood
in terms of a non-cooperative network being established at $y \sim 0.16$; this network
leads to an enhanced  conduction since hopping does not generate unfavorable distortions.

\begin{figure}[t]
\includegraphics[width=0.55\textwidth]{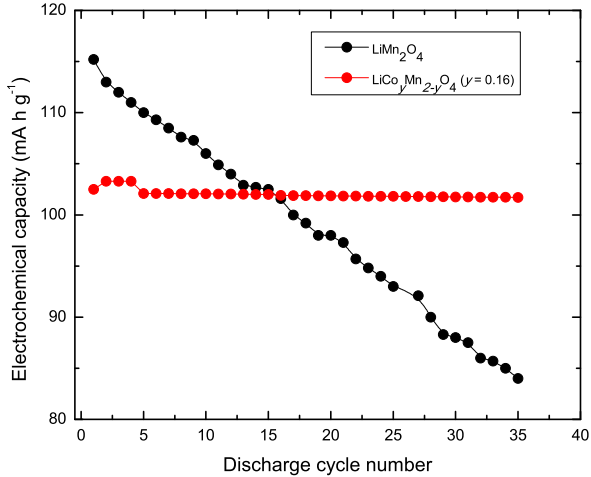}
\caption{Electrochemical capacity versus discharge cycle number (adapted from Ref. \onlinecite{rag2}).
Capacity fades with repeated cycling  for the undoped cathode whereas it
remains unchanged at $y \sim 0.16$. See Fig. \ref{Fig. 6}
 for corroborative doping dependence of capacity.}\label{Fig. 4}
\end{figure}

\begin{figure}[b]
\includegraphics[width=0.55\textwidth]{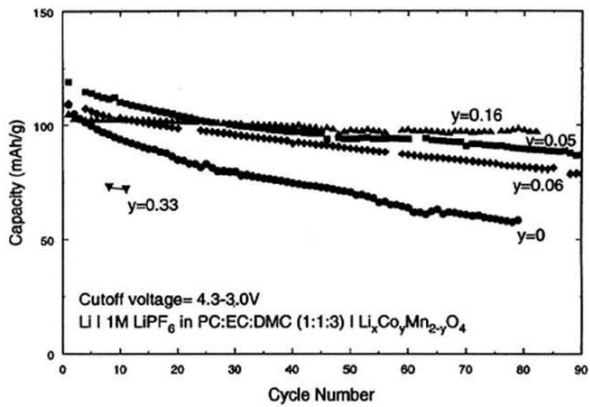}
\vspace{-3.0cm}
\caption{Variation of the capacity as a function of cycle number for
Li/LiCo$_y$Mn$_{2-y}$0$_4$ cells at various cobalt dopings.
At higher doping $y \sim 0.16$, capacity remains unchanged after repeated cycling.
Reproduced from Ref. \onlinecite{arora}.
}\label{Fig. 6}
\end{figure}

Finally, we will discuss capacity fading as displayed in Figs. \ref{Fig. 4} and \ref{Fig. 6}. In the undoped
case and at lower dopings (i.e., $y < 0.16$), the network of cooperative/coherent normal-mode distortions  restricts  lithiation (delithiation) of the cathode
material Li$_{\rm x}$Mn$_2$O$_4$; consequently, each time  only a fraction of  the un-lithiated (lithiated)
material gets lithiated (delithiated). On increasing the doping to $y \gtrsim 0.16$,
a network of non-cooperative normal-mode distortions is established which facilitates both the lithiation
and the delithiation processes. Thus, while there is capacity fading upon repeated cycling
at lower values of doping (i.e., $y < 0.16$), the capacity remains constant
for $y \gtrsim 0.16$. However, for $y \gtrsim 0.16$,  at higher doping values the capacity is less due to
decrease in the number of carriers. Thus ideally, it is best to use
LiCo$_y$Mn$_{2-y}$O$_4$ at $y \sim 0.16$ for optimal electrochemical performance.

\section{conclusions}
From the results of the thermal expansion and electrical conductivity measurements on LiCo$_y$Mn$_{2-y}$O$_4$ spinel,
within the framework of coherent normal-mode  physics,
we provide an explanation for the optimum doping level that generates best capacity performance in
Li/LiCo$_y$Mn$_{2-y}$O$_4$ cells. At $y =0$, the material shows a first-order phase transition attributed to cooperative normal-mode distortions. These cooperative distortions lower the electrical conductivity as well as the efficiency of Li insertion and de-insertion into the spinel structure, leading to lowering of discharge capacity with cycle number. With increase in the doping `$y$', the cooperative-normal-mode network starts to weaken, and at $y \sim 0.16$ the cooperative network is completely broken. Any further increase in `$y$', only causes a lowering of the capacity due to a decrease in the number of charge carriers. The  $y \sim 0.16$ is thus the ideal doping level for realizing stable discharge capacity of the battery.

\section{Acknowledgements}
S. Y. acknowledges stimulating discussions with P. B. Littlewood, M. M. Thackeray, H. Iddir, and J. W. Freeland.
The authors thank A. Ghosh for help with the figures and the students of P. Mandal for help with the
measurements. 

\end{document}